# Structure and Synthesis of graphene oxide

**Sun, Ling**[*]

*Beijing Guyue New Materials Research Institute,* College of Material Science & Engineering, Beijing University of Technology

Pingleyuan 100, Chaoyang district, Beijing 100124

**Abstract**

Graphene oxide (GO) is regarded as one typical two-dimension structured oxygenated planar molecular material. Researchers across multiple disciplines have paid enormous attention to it due to unique physiochemical properties. However, models used to describe the structure of GO are still in argument and ongoing to update. Currently, synthesis methods for massive production are seemingly abundant but in fact, dominated by few thought systems. We herein aim to give a mini but critical review over the synthesis of graphene oxide as well as its structure, involving relative peer work.

---

[*] Corresponding author. E-mail: sunling@bjut.edu.cn (Ling Sun)

# 1. Introduction

Graphene oxide (GO) is the oxidized analogy of graphene, recognized as the only intermediate or precursor for obtaining the latter in large scale, [1] since the English chemist, sir Brodie first reported about the oxidation of graphite centuries ago [2]. About thirty years ago, the term *graphene* was officially claimed to define the single atom-thin carbon layer of graphite [3], which structurally comprises sp$^2$ hybridized carbon atoms arranged in a honeycomb lattice, rendering itself large surface area and some promising properties in terms of mechanical, electrical, and others. [4, 5] Despite these extraordinary properties, purely single-layer graphene remains very limited success in practical applications due to the difficulties in the large-scale formation of specifically organized structures. [6] But the precursor GO has advanced much in both academics and industries in the last decades because of its readiness by exfoliating bulk graphite oxide facilely prepared from the oxidation of graphite. [7, 8] This bottom-down chemical strategy features the upmost flexibility and effectiveness thereby arousing the greatest interest in practical applications.

Now it is seemingly clear that GO is a non-stoichiometric chemical compound of carbon, oxygen, and hydrogen in variable ratios which largely yet partly depend on the processing methodologies. [2, 9–11] GO possesses abundant oxygen functional groups that are introduced to the flat carbon grid during chemical exfoliation, evidenced as oxygen epoxide groups (bridging oxygen atoms), carbonyl (C=O), hydroxyl (-OH), phenol, and even organosulfate groups (impurity of Sulphur). [12, 13]In other word, these defects of various kinds are brought into the naturally intact graphene structure, further categorized into on-plane functionalization defects and in-plane lattice defects (vacancy defects and hole defects) which are semi-randomly distributed in GO's σ-framework of the hexagonal lattice. [1, 7] Such a defect-rich structure leads to a set of unique properties of GO and render it availability and scalability of consequent applications via post treatments, e.g. chemically-derived graphene-like materials, functionalized graphene-based polymer composites, sensors, photovoltaics, membranes[14] and purification materials. On the detailed structure of GO, however, it remains some ambiguous and literature reports are still in argument (Fig.1). [11, 13, 15–23] In addition, methods in regard to synthesis of GO has been massively researched in the past few years. The effectiveness and environmental benignity was core driven force for the continuous evolvement. Herein, we update the progress and make a short yet critical review on model structures of GO as well as the synthesis.

## 2. Graphene oxide structure

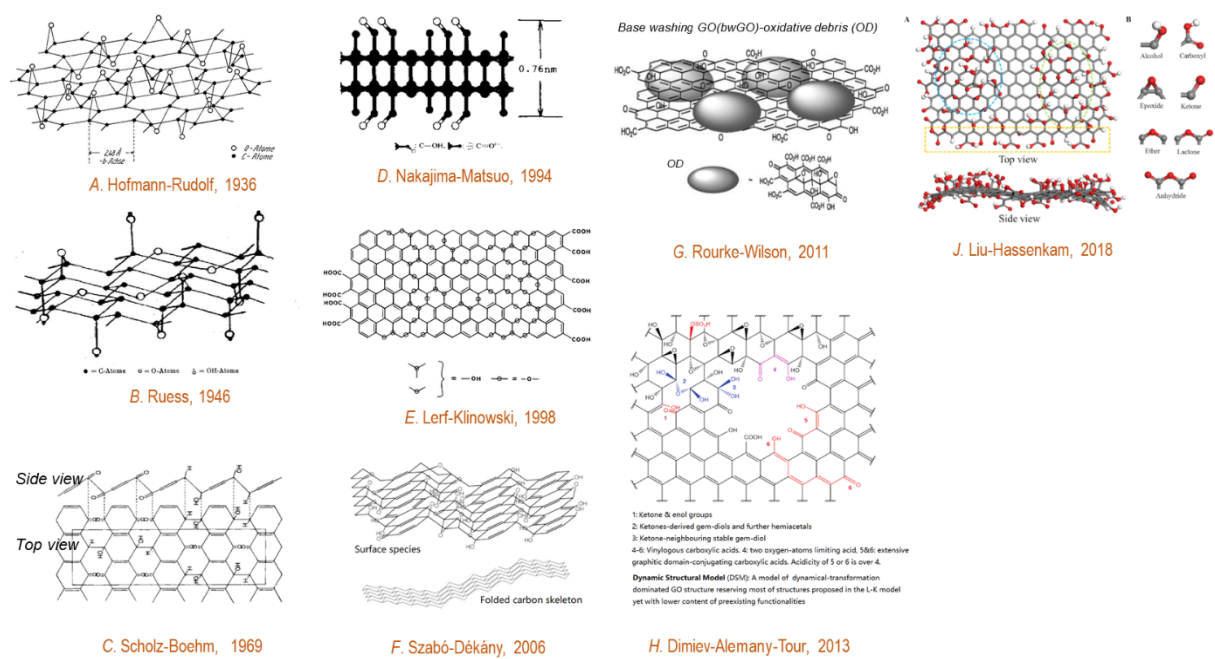

Figure 1. Theoretical models for structures of graphite or graphene oxide.

In 1936, Hofmann and Rudolf [22] proposed the first GO structure (Figure 1, the top-left 1st schematic structure) in which a deal of epoxy groups are randomly distributed on the graphite layer, and then in 1946, Ruess [21] updated the Hofmann model by incorporation with hydroxyl entities and alternation of the basal plane structure ($sp^2$ hybridized model) with a $sp^3$ hybridized carbon system. By contrast, in 1969, Scholz and Boehm [19] proposed a less ordered structure with C=C double bonds and periodically cleaved C-C bonds within the corrugated carbon layers and hydroxyl, carbonyl groups in different surroundings, free from ether oxygen. More further in 1994, Nakajima and Matsuo [18] proposed a stage 2 graphite intercalation compound (GIC)-resembled lattice framework based on the fact that fluorination of graphite oxide gives the same X-ray diffraction pattern as that of stage 2-type graphite fluoride, (C2F)n. In 1998, Lerf and Klinowski et al. [17] characterized their GO by the $^{13}C$ and $^1H$ nuclear magnetic resonance (NMR), and subsequently found the 60 ppm line better related to epoxide groups (1,2-ethers) other than 1,3 ethers, and the 130 ppm line to aromatic entities and conjugated double bonds. The carbon atoms attached to OH groups slightly distorted their tetrahedral structure, resulting in partial wrinkling of the layers. Accordingly, they proposed a model featuring a nearly-flat carbon grid structure with randomly distributed aromatic regions with unoxidized benzene rings and regions with aliphatic six-membered rings. However, all these earlier models could not well explain the origin of the planar acidity of GO, which is now a well-understood chemical property for GO. Thereafter, Szabó and coworkers in 2006 [16] revived but a little modified the Scholz-Boehm model, etc. by again examining the results from elemental analysis, transmission electron microscopy, X-ray diffraction, diffuse reflectance infrared Fourier transform spectroscopy, X-ray photoelectron spectroscopy, and electron spin resonance besides NMR. They then proposed a carboxylic acid-free model comprising two distinct domains: trans-linked cyclohexyl species interspersed with tertiary alcohols and 1,3-ethers, and a keto/quinoidal species corrugated network. Interestingly, as to the phenomenon of GO in basic solution Roukre et al. [23] found that GO decomposed into slightly oxygenated graphene part and strongly graphene-bound oxidative debris (OD) upon suffering a base washing, and then suggested a simple OD-base washed GO two-component model, which was much different from those previously proposed, upgrading the way we used to understand about GO. Besides, they also mentioned about the metastability of unwashed GO, which reminded us of previous room-temperature metastable GO film [24], while the internal mechanism of external stimuli-responded structural instability was lack of sufficient investigation. In 2013, Dimiev et al. [11] revisited the structure via acid titration and ion exchange experiment in term of acidity of GO and proposed a novel dynamical structural mode (DSM), which describes the evolution of several carbon structures with attached water beyond the static Lerf's model. More recently, Liu et al. [25] experimentally observed oxygen bonding and evidenced the C=O bonds on the edge and plane of GO, confirming parts of earlier proposed models, especially the L-K model.

Amongst these models from 1936 through 2018, the L-K model has been the most widely used due to the good interpretability over the majority of experimental observation, and easiness of further adaption/modification, for example, with which as the starting basis the Rourke-Wilson model [23], Dimie-Alemany-Tour model [11], etc. were successively publicized and continually paved the way forwards. Nonetheless, the unique two dimensional geometry of GO, has been universally accepted as the basic character, and this laid a key foundation for GO subsequently blooming in enormous researches, especially after the Nobel Award honored the discovery of graphene in 2010.

## 3. Synthesis and Progress: Solution-processed Methods

Table. 1 Methods for preparation of GO

| Methods | Carbon Source | Oxidants | Reaction Time for graphite oxide | Temperature ℃ | Features |
|---|---|---|---|---|---|
| Brodie,1859[2] | graphite | $KClO_3$, $HNO_3$ | 3~4 d | 60 | earliest method |
| Staudenmaier,1898[26] | graphite | $KClO_3$, $HNO_3$, $H_2SO_4$ | 96 h | RT | improved efficiency |
| Hummers,1958[10] | graphite,~44μm | $KMnO_4$, $NaNO_3$, $H_2SO_4$ | <2 h | <20–35-98 | water-free, less than 2hrs processing |
| Fu,2005[27], | graphite | $KMnO_4$, $NaNO_3$, $H_2SO_4$ | <2 h | 35 | validate $NaNO_3$ unnecessary |
| Shen,2009[28] | graphite colloidal ~10μm | Benzoyl peroxide(BPO) | 10 min | 110 | fast and non-acid |
| Su,2009[29] | sonicated graphite, <3000μm | $KMnO_4$, $H_2SO_4$ | 4h | RT | large-size GO |
| Marcano,2010&2018[9] | graphite ~150μm | $H_2SO_4$, $H_3PO_4$, $KMnO_4$ | 12 h | 50 | bi-component acids, high yield |
| Sun,2013[30] | expanded graphite | $KMnO_4$, $H_2SO_4$ | 1.5h | RT-90 | size-confined high yield, safe |
| Eigler,2013[31] | graphite, ~300μm | $KMnO_4$, $NaNO_3$, $H_2SO_4$ | 16 h | 10 | high-quality GO |
| Chen,2015[32] | graphite, 3-20μm | $KMnO_4$, $H_2SO_4$ | <1h | <20-40-95 | high-yield |
| Panwar,2015[33] | graphite | $H_2SO_4$, $H_3PO_4$, $KMnO_4$, $HNO_3$ | 3h | 50 | tri-component acids, high yield |
| Peng,2015[34] | graphite >10μm | $K_2FeO_4$, $H_2SO_4$ | 1h | RT | high-yield, less pollution |
| Rosillo-Lopez,2016[35] | defective arc-discharge carbon | $HNO_3$ | 20 h | RT | nano-sized GO |
| Yu,2016[36] | graphite,~44μm | $K_2FeO_4$, $KMnO_4$, $H_2SO_4$, $H_3BO_3$ | 5h | <5-35-95 | less manganite impurity, less acid, high yield |
| Dimiev,2016[37] | graphite | $(NH_4)_2S_2O_8$, 98%$H_2SO_4$, fuming $H_2SO_4$ | 3~4h | RT | lightly oxidized, 25nm thick,~100% conversion |

| Pei,2018[38] | graphite foil | H$_2$SO$_4$ | <5min | RT | electrochemistry support; high efficiency and high yield |
| Ranjan,2018[39] | graphite | H$_2$SO$_4$, H$_3$PO$_4$, KMnO$_4$ | >24h | <RT-35-95 | Cool the exothermal reaction to keep safe |

RT: abbreviation of room temperature.

As mentioned above, Brodie [2] reported the change of graphite blending with strong oxidants, and this can be regarded as the earliest preparation of GO, although he termed the final material as "graphic acid", which we know now as graphite oxide. As time has gone, chemistry of graphite oxide has advanced much with efforts of scientists worldwide, especially when graphite oxide is known capable of transformation into graphene oxide/graphene as easily-obtainable precursor. Graphite oxide-derived GO becomes one most tangible outcome of the graphene research worldwide in terms of large scale production and commercialization prospects. The top-down strategy endows the preparation with visible flexibility and relatively low cost of input. As a result, methods adopting graphite including that expanded form as starting material have been widely used in laboratory researches and industrial production/applications (Table 1).

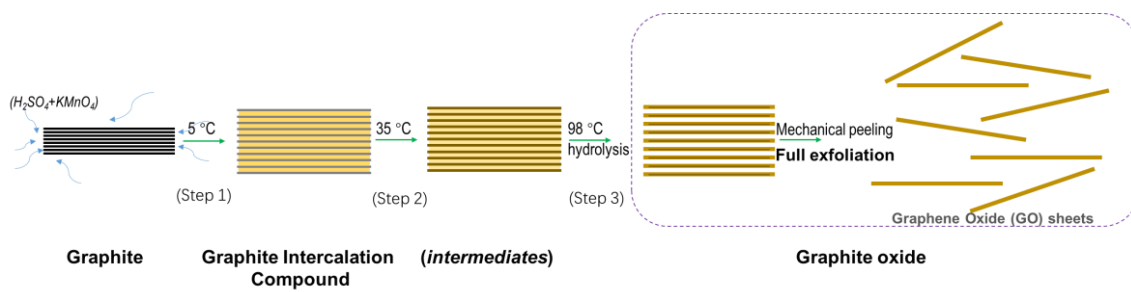

Figure 2. Schematic diagram of GO preparation via Hummers-Offeman method [10]

Nowadays, the Hummers method of 1958 is still widely employed to delaminate and oxidize graphite because of the improved convenience in comparison with that of Brodie and earlier followers. Abstracted from original method, it relies much on a mixture of sulfuric acid and potassium permanganate (Figure 2, first two steps) and the whole procedure comprises three steps: a period for the intercalation of graphite and a simultaneous/subsequent oxidization of the above-mentioned graphite intercalation compounds (GICs); next, to obtain homogeneous GO solution (Figure 2, the third step), graphite oxide is then hydrolyzed and straightforward exfoliated into single sheets via mechanical peeling, like sonication[40], swirling (by shearing stress)[41], or others. Noticeably, many groups prefer using a some-time certain-strength ultra-sonication to completely break up the stacked structure of graphite oxide into GO sheets.

As listed in the Table 1, prior to the Hummers' work, Brodie [2] provided a oxidization mixture recipe of fuming HNO$_3$ and KClO$_3$ in 1859, but clearly the method was tedious and not benign. Then Staudenmaier [26] further adjusted the acid component via addition of H$_2$SO$_4$, rendering the method with fast completion via one single-vessel reaction and thereby improved the processing yield. However, it remained large space for further cutting off the processing time and reducing the outcome of hazards, such as toxic gases. Years late, did Hummers and Offeman [10] express the attitude to these early methods "described in the literature is time consuming and hazardous", so they themselves altered the oxidant to a water-free mixture via replacing the $HNO_3$, $KClO_3$ with $KMnO_4$. Consequently, new process taking less than two hours with lower working temperature turned out to be more effective and less hazardous than before. Towards large-scale and safe production of GO, it remained to some extent unsatisfactory problems. Consequently, variations were proposed in succession and seem to reach a research summit appearing in these years (2013-2018).

Again, the Hummers method stresses on the three-stage reaction [27]: low-temperature (below 5 ℃) intercalation, mid-temperature (~35 ℃) oxidizing of the GIC and high-temperature (98 ℃) hydrolysis of consequences. A Chinese team led by Fu [27] further investigated each stage and compared the effect of parameters' change concerning e.g., the mass ratios amongst graphite, $H_2SO_4$ and $KMnO_4$, and the ways to add water. They concluded the $NaNO_3$ did not play important role in adequate oxidization of graphite, and suggested the cancellation of use of $NaNO_3$. To the best of our knowledge, this probably is the earliest research to clarify the redundancy of $NaNO_3$ in Hummers method. It becomes understandable that these changes not only simplify the process and the composition of discharged water, but also alleviate the evolution of toxic gasses, e.g., $NO_2/N_2O_4$. These findings were also re-declared or replicated in other respective researches by Tour group (2010) [9], Fugetsu group(2013) [30], Shi group (2013) [42] and others [29, 33, 36, 37, 39].

One other point we highlight is the widely accepted ingredient, the concentric $H_2SO_4$. It features the readiness in use due to the relatively high boiling point, non-volatility, and low cost. Therefore, it was retained in most variations to original Hummers method [9, 29, 31, 33, 34, 36, 39], except those changed either the oxidization phase [28] or the raw material [35]. The amount of acid per gram graphite consumes so large (>10ml 95~98% purity $H_2SO_4$ per gram of graphite) that we have to take serious the recycle use of acid and the avoidance of accidental leakage into the environment. In other words, the post-treatment for the GO purification should have been paid more attention in. Consequently, considering the disposing cost, it is understandable the overall cost per gram GO obtained remains as high as that of the skyscraper. This is undoubtedly hindered the prosperity of GO applications. However, novel methods with less acid are scarcely reported.

Sun et al.[30] conceived a simple principle as shown in Figure 3. They hypothesized that graphite in varying sizes and textures carried out the oxidations consuming different time: under an identical condition, the intercalation over larger graphite is longer than that of smaller one, the tightly-structured graphite is likewise longer than the loosely-structured (Figure 3A). By this implication, they selected commodity expanded graphite in different sizes. They observed the mixture underwent a fast and distinctly volumetric expansion companied with a continuous magnetic stirring, and formed a pale-gray foam-like slurry in the end (Figure 3B).[30] The addition of water therefore became really simple and secure without fearing the splashing of acid. Due to the volumetric expansion as the visible indicator conveniently reflected the end of a reaction and improved security, one modified Hummers method with expanded graphite was also proposed. Not only a lower demand for acid worked (10 ml versus 13 ml per gram of graphite before), but also nearly one hundred percent yield with smaller graphite ($D_{50}$ ~15 μm) suggested available in contrast to larger one ($D_{50}$ ~50 μm). Coincidently, Chen et al.[32]

found the flake graphite with sizes in the range of 3–20 μm be completely converted into GO without additional centrifugation, yet with a routine mass configuration.

Taken together, the size confinement is doomed to further differentiate current strategies, in which the total acid demand for smaller graphite (< 20 μm) might decrease even more, e.g., by one fifth or one fourth; for the larger graphite, a trade-off between the acid and oxidants needs further searching.

For the latter, there has been a few advances. Their involved mechanisms were by either extending the time of intercalation/oxidation reaction or changing the oxidants. Huang et al. [43] introduced one-pot chemical oxidation method by simply stirring large graphite (~500 μm) in a mixture of acids and potassium permanganate at room temperature for 3×24 hours, achieving large-area GO sheets with nearly 100% conversion. Similarly, Eigler et al. [31] prolonged the low-temperature oxidation of graphite (~300 μm) over 16 hours and then stepwise fed diluted sulfuric acid and water for further hydrolysis. Noting that the entire process was at a temperature below 10 ℃. As a result, a new form of GO was prepared consisting of an hexagonal intact σ-framework of C-atoms, which was easily reduced to graphene that is no longer dominated by defects.

In contrast to the time variation, Peng et al. reported a $K_2FeO_4$-based modified hummers method (Figure 4A), replacing $KMnO_4$ with $K_2FeO_4$ due to the higher oxidation potential. [34] Such iron-based oxidization realized a green production of GO in 1 h and enabled the recycling of sulfuric acid and eliminating the emission of heavy metals and toxic gas, as forwarded from the paper. This work ignited the ferric acid-based applications in modifying methods targeting high greenness and conversion efficiency. However, the aforementioned method did not go well with the repeatability, due to the instability of iron when dissolving in acidic aqueous. [44] Even though the difficulty exists, iron-based methods cannot stop their steps. Yu et al. [36] used $K_2FeO_4$ to partly replace $KMnO_4$ (Figure 4B) and at the same time reduced the acid amount to the same extent as previously Sun et al. reported in their modified $NaNO_3$-free method, demonstrating well effectiveness except for a longer time versus original Hummers method.

Ultrasound is also capable of reducing both the vertical and lateral size of graphite, so several modifications included the pre-sonication or synchronous sonication over the graphite [35, 45]. Rosillo-Lopez et al.[35] pre-sonicated arc-discharge carbon source and then oxidized them in half-concentrated nitric acid to successfully obtain the nano GO. Yang et al. [45] proposed their method by taking advantage of the synergistic effect between intercalation and sonication, resulting in a strong decrease in demand for time and acid as compared to that of the Hummers method. Likewise, pre-oxidization of graphite is also beneficial for synthesis. It could enlarge the interlayer spacing, in other words, decreasing the vertical size of graphite, therefore, alleviating the resistance for molecules/ions intercalation into interlayers. Following this strategy, Kovtyukhova et al. [46] found incompletely oxidized graphite-core/graphite oxide-shell particles always existed. In 1999, they tried a pre-oxidation with a tri-component mixture $H_2SO4-K_2S_2O_8-P_2O_5$ over graphite followed by the Hummers method, which succeeded in complete oxidation of graphite. So far, this methodology remains inspiring for subsequent research. All the above methods could be categorized into such a liquid-phase chemical method. And some of them have been used for moderate-scale production of commercial GO in spite of high time cost and potential environmental risk.

We also noticed some other remarkable progress had been made: (1) By solution electrochemical exfoliation. In detail, Pei et al. [38] employed a lower-voltage (DC 1.6V) power to drive sulfate radical into graphite foil to generate intercalated GICs in 98% sulfuric acid and a higher voltage (DC 5V) to oxide the obtained GICs into graphite oxide. This method was found based on a mechanism of water electrolytic oxidation towards graphite. Although featuring fast and high yield, rather greener that sole liquid-phase chemical oxidation, this method still suffers from the critical use of strong high and concentrated acids. And for industrial scale, a set of specific apparatus are necessary since it cannot work without electric power. (2) By one single-step exfoliation. Shen et al. [28] discovered at high-temperature (~110 °C) molten organic oxidant benzoyl peroxide can fast intercalate and oxidize graphite to form GICs and then through sonication and wash, GICs were exfoliated into GO sheets; Dimiev et al. [37] developed an absolutely water-free tri-component strong acidic system to fast split graphite, of which the expansion ended with the formation of a greenish-yellow foam. Although what they reported was much likely graphene, on graphene were still oxygen-containing functionalities. The novelty of these methods would aspire the community of continuous insistence [47, 48], while the high risk hinders their way into large scale production.

With the summary, we clearly sense the present synthesis much depends on the intercalation chemistry of graphite in terms of sizes, the time, and/or the kinds of oxidants. Even though less suffering from explosion risk, and environmental pollution along by the efforts from chemists, all these methods remain intrinsic limitations associated with acid recycle and post water treatment. However, we keep ourselves with great confidence that these issues can be step-by-step addressed by, for example, using relatively safer yet highly efficient oxidizing intercalating agents, applying electrochemical oxidation or other smarter ways. Besides, the expanded graphite as starting materials could be another way for the turning-around, due to commercial availability, pre-oxidized feature and its potential to further decrease the dose of oxidants,

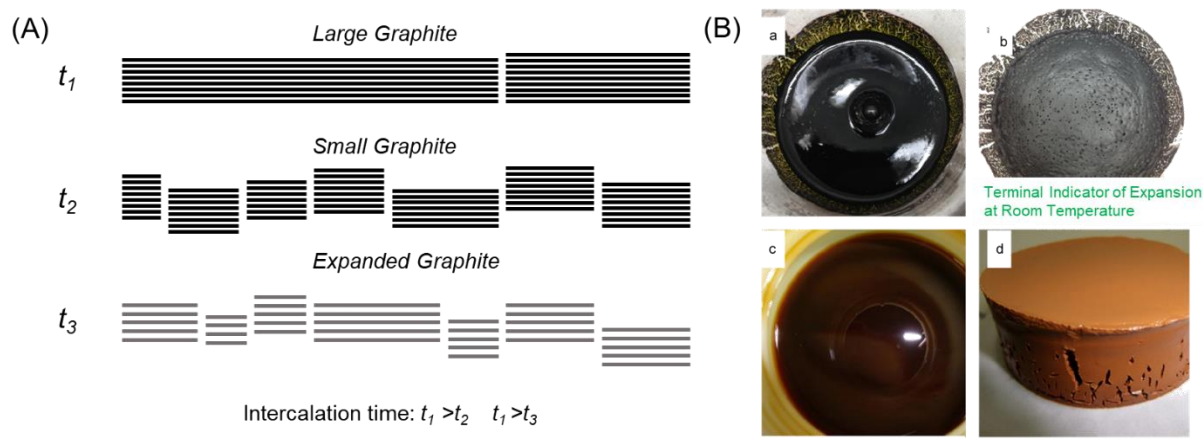

Figure 3. (A) Difference in intercalation time for graphite with different lateral sizes or vertical textures, adapted from the Sun (2013) paper [30] ; (B) Photos related to the preparation procedure of GO from expanded graphite ~15 μm following the Sun-Fugetsu modified Hummers method: (a) mixing, (b) the form-like slurry after full expansion ends, (c) a light-brown GO "cake" after high-temperature hydrolysis, (d) a stock solution of GO after purification.

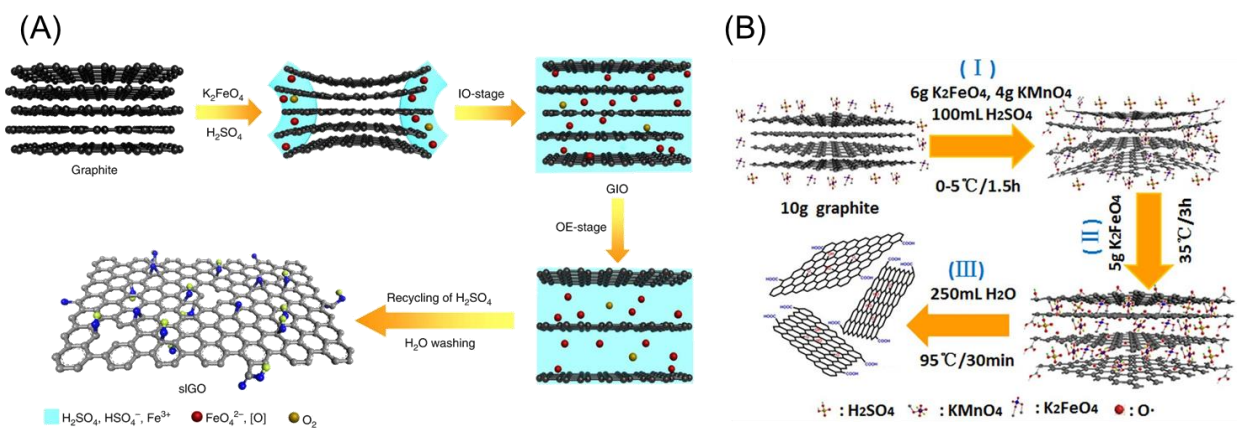

Figure 4. Iron-based modified methods for the preparation of GO. (A). Mechanism of GO synthesis with the oxidant of $K_2FeO_4$ following the Peng-Gao modified method [34]; (B). Mechanism of GO synthesis with the $K_2FeO_4$-$KMnO_4$ bi-component oxidant following the Yu-Zhang modified method [36].

## 4. Prospective

From static to dynamic model, with theoretical speculation to experimental observation, the time dependence has been taken in account for intrinsic property investigation of GO. Fascinating measures and characterizations are in continuity on the way to development. Therefore, it is expected that complete characters of the GO shall become fully unraveled and universally acceptable, therefore beneficial for understanding and guiding in-depth applications.

Research on the synthesis of GO now is at such a plateau where the Hummers method (in relation to concentrated sulfuric acid) dominates as the mainstream strategy. Greener and more effective breakthroughs are much desirable, calling more contributions from more intelligence with multi-discipline field background. Moreover, real synthesis of GO has to tackle the problems in relation to the uniformity of sizes and properties. It is unrealistic for downstream industries to embrace GO products from upstream entities having varying performances.

**Acknowledgments**

This work is partly financially supported by Beijing university of technology (105000546317502,105000514116002) and Beijing Municipal Education Commission (KM201910005007)